\documentclass[twocolumn,english,aps,prb,twocolum,superscriptaddress,bibnotes,amsmath,amssymb,floatfix]{revtex4-2}

\usepackage[colorlinks=true,citecolor=blue,linkcolor=magenta]{hyperref}
\usepackage{changes}
\usepackage{lineno}
\usepackage{stmaryrd}
\usepackage{bm}
\usepackage{graphicx}
\usepackage{epsfig}
\usepackage{color}
\usepackage{amssymb}
\usepackage{amsmath}
\usepackage{array}
\usepackage[loose]{units}
\usepackage{braket}
\usepackage{soul}
\usepackage{hyperref}
\usepackage{cleveref}
\usepackage[caption=false]{subfig}
\usepackage[english]{babel}
\usepackage{letltxmacro}
\usepackage{siunitx}
\usepackage{dsfont}
\usepackage{ulem}

\LetLtxMacro{\ORIGselectlanguage}{\selectlanguage}
\makeatletter
\DeclareRobustCommand{\selectlanguage}[1]{%
    \@ifundefined{alias@\string#1}
      {\ORIGselectlanguage{#1}}
      {\begingroup\edef\x{\endgroup
         \noexpand\ORIGselectlanguage{\@nameuse{alias@#1}}}\x}%
}

\usepackage{enumitem}
\setitemize{noitemsep,topsep=0pt,parsep=0pt,partopsep=0pt}

\newcommand{\MPINAT}{Max Planck Institute for Multidisciplinary Sciences, D-37077 G\"{o}ttingen, Germany}
\newcommand{\GOE}{Georg-August-Universit\"{a}t G\"{o}ttingen, D-37077 G\"{o}ttingen, Germany}
\newcommand{\EPFL}{Swiss Federal Institute of Technology Lausanne (EPFL), CH-1015 Lausanne, Switzerland}
\newcommand{\CfQS}{Center for Quantum Science and Engineering, EPFL, Lausanne, Switzerland}

\begin{document}

\title{Cavity-mediated electron-photon pairs}

\author{Armin Feist}
\thanks{These authors contributed equally.}
\affiliation{\MPINAT}
\affiliation{\GOE}
\author{Guanhao Huang}
\thanks{These authors contributed equally.}
\affiliation{\EPFL}
\affiliation{\CfQS}
\author{Germaine Arend}
\thanks{These authors contributed equally.}
\affiliation{\MPINAT}
\affiliation{\GOE}
\author{Yujia Yang}
\thanks{These authors contributed equally.}
\affiliation{\EPFL}
\affiliation{\CfQS}
\author{Jan-Wilke Henke}
\affiliation{\MPINAT}
\affiliation{\GOE}
\author{Arslan Sajid Raja}
\affiliation{\EPFL}
\affiliation{\CfQS}
\author{F. Jasmin Kappert}
\affiliation{\MPINAT}
\affiliation{\GOE}
\author{Rui Ning Wang}
\affiliation{\EPFL}
\affiliation{\CfQS}
\author{Hugo Louren\c{c}o-Martins}
\affiliation{\MPINAT}
\affiliation{\GOE}
\author{Zheru Qiu}
\affiliation{\EPFL}
\affiliation{\CfQS}
\author{Junqiu Liu}
\affiliation{\EPFL}
\affiliation{\CfQS}
\author{Ofer Kfir}
\affiliation{\MPINAT}
\affiliation{\GOE}
\author{Tobias J. Kippenberg}
\email{tobias.kippenberg@epfl.ch}
\affiliation{\EPFL}
\affiliation{\CfQS}
\author{Claus Ropers}
\email{claus.ropers@mpinat.mpg.de}
\affiliation{\MPINAT}
\affiliation{\GOE}

\maketitle
\date{\today}

\section*{Abstract}
\textbf{Advancing quantum information, communication and sensing relies on the generation and control of quantum correlations in complementary degrees of freedom. Here, we demonstrate the preparation of electron-photon pair states using the phase-matched interaction of free electrons with the evanescent vacuum field of a photonic-chip-based optical microresonator. Spontaneous inelastic scattering produces intracavity photons coincident with energy-shifted electrons. Harnessing these pairs for correlation-enhanced imaging, we achieve a two-orders of magnitude contrast improvement in cavity-mode mapping by coincidence-gated electron spectroscopy. This parametric pair-state preparation will underpin the future development of free-electron quantum optics, providing a pathway to quantum-enhanced imaging, electron-photon entanglement, and heralded single-electron and Fock-state photon sources.}\\

\section*{Introduction}
Optical parametric processes generate quantum correlations of photons, without changing the state of the optical medium involved. Entangled photons from parametric down conversion~\cite{Burnham1970} are an essential resource for quantum communication~\cite{Gisin2007}, heralded single photon sources~\cite{Eisaman2011, Signorini2020}, and quantum teleportation~\cite{Bouwmeester1997}. Over the past decade, such ``twin beam" pairs have been extended to other physical contexts, including photon-phonon correlations~\cite{TarragoVelez2020}, non-classical states~\cite{Hong2017} and entanglement of micro-mechanical systems~\cite{Riedinger2018}.
Free-electron beams are an emerging target for quantum manipulation and sensing, promising quantum-enhanced imaging~\cite{Kruit2016, Juffmann2017, Turner2021, Li2018, Rotunno2021}, spectroscopy~\cite{Priebe2017, Kfir2019, DiGiulio2019, Pan2019, BenHayun2021, Kfir2021,Tsarev2021}, and excitation~\cite{Pan2020, Yalunin2021, Zhao2021, Ratzel2021}. A variety of technologies bridging electron microscopy and photonics~\cite{Polman2019, GarciadeAbajo2021_review} are being established to join the most powerful probes in atomic-scale imaging and spectroscopy, respectively. Among these, stimulated near-field scattering~\cite{Barwick2009, GarciadeAbajo2010a, Feist2015} offers mode-specific probing of optical properties~\cite{Yurtsever2012, Piazza2015, Wang2020_cavity, Kurman2021, Liebtrau2021,Henke2021}, whereas spontaneous electron energy loss and cathodoluminescence access electronic transitions and the total photonic density of states~\cite{Kociak2017, Polman2019, GarciadeAbajo2021_review}. Structural design has been shown to allow for a tailoring of the spectral and spatial properties of electron-driven radiation~\cite{Talebi2014, Saito2015, Remez2017, Mignuzzi2018, Roques-Carmes2019, Christopher2020}.
Harnessing quantum optics approaches, measurements of photon statistics have been employed to reveal single quantum emitters~\cite{Bourrellier2016} or photon bunching~\cite{Meuret2015, Sola-Garcia2021, Scheucher2021}. Theoretical work predicted single-photon cathodoluminescence into a fiber waveguide~\cite{Bendana2011}, and recent experiments studied the electron-induced excitation of whispering gallery modes~\cite{Muller2021,Auad2022} and optical fibers~\cite{Scheucher2021, Uemura2021}. However, impeded by a lack of mode-specific and sufficiently strong coupling, correlations between electrons and well-defined photonic states have remained elusive.
In this work, we employ spontaneous inelastic scattering via the evanescent field of a high-Q photonic-chip-based optical microresonator to generate free-electron cavity-photon pair states. We characterize the dual-particle heralding efficiencies and demonstrate that coincidence imaging of the cavity mode yields a drastic background suppression compared to mapping with either electrons or photons alone.

\section*{Electron-driven generation of cavity photons}
\begin{figure} [!t]
\centering
\includegraphics[width=0.5\textwidth]{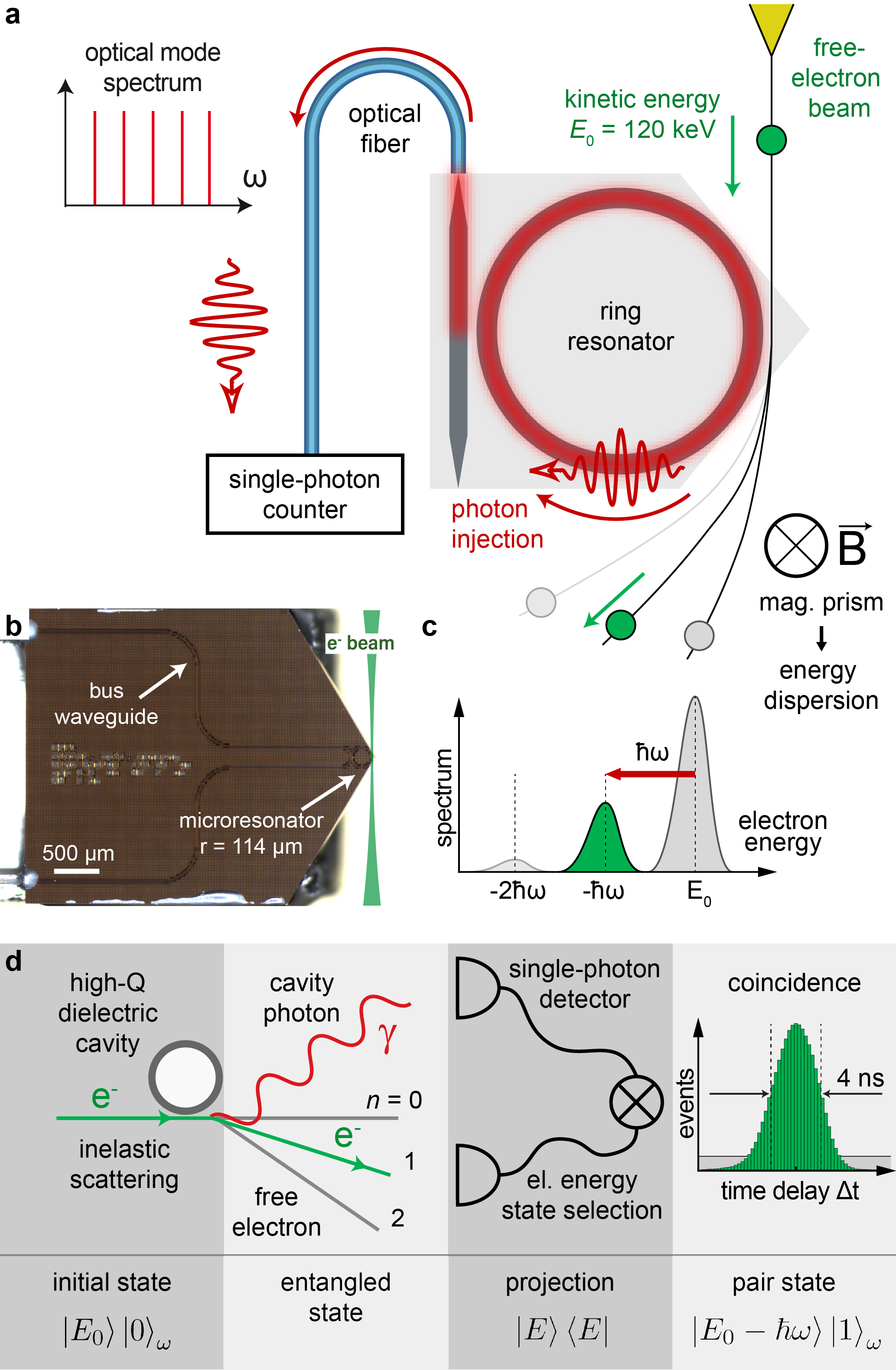}
\caption{\textbf{Coincidence detection of electron-photon pairs via a high-Q integrated-photonic microresonator. a)} Electrons (green) extracted from a field emission tip (yellow) traverse a fiber-coupled microring resonator at 120-keV kinetic energy. The electrons spontaneously generate intracavity photons detected with single-photon sensitivity. Relative timings and energy losses of the electrons are analyzed with an event-based detector behind a magnetic prism (\textbf{c}). \textbf{b)} Optical microscope image of the photonic chip with the bus waveguide and the Si\textsubscript{3}N\textsubscript{4} microresonator. The green path indicates the trajectory of the electron beam (not to scale) passing the microresonator parallel to the chip surface. \textbf{d)} Illustration of the cavity-mediated inelastic electron-photon scattering process and coincidence detection scheme. Electron-photon correlations demonstrate the generation of single photons with corresponding energy lowering of the electrons by $\hbar\omega$ (green curves in \textbf{c} and \textbf{d}, 4-ns timing accuracy).
}\label{fig_setup}
\end{figure}

The interaction of electron beams with cavities and resonant structures represents a universal scheme for generating electromagnetic radiation. In the quantum optical description, the inelastic scattering can be modeled as a coupling of free electrons to optical vacuum fields (cf. Fig.~\ref{fig_setup})~\cite{Asenjo-Garcia2013}.
Scattering with the evanescent field of the optical microresonator (cf. Fig.~\ref{fig_setup}a), an electron at energy $E$ generates intracavity photons at frequencies $\omega$ in an energy-conserving manner, described by the scattering matrix $\hat{S} = \exp\left(g_{\mathrm{qu}} \hat{a}^\dag \hat{b} - h.c.\right)$, where $\hat{a}^\dag$ is the creation operator of the optical mode, $\hat{b}$ is the electron-energy lowering operator and $g_{\mathrm{qu}}$ is the vacuum coupling strength~\cite{Feist2015}.
The interaction induces entanglement between the electron energy and the cavity population, and results in the state
\begin{gather}
    |\psi_\mathrm{e}, \psi_\mathrm{ph}\rangle =\sum_{n = 0}^\infty c_n|E - n\hbar\omega\rangle|n\rangle,
\end{gather}
with the coefficients $c_n=\exp(-\frac{|g_{\mathrm{qu}}|^2}{2})\frac{g_{\mathrm{qu}}^n}{\sqrt{n!}}$ corresponding to Poissonian scattering probabilities~\cite{ Kfir2019,DiGiulio2019}. For a weak vacuum coupling strength $|g_{\mathrm{qu}}|\ll 1$, the state is dominated by the zero- and one-photon contributions:
\begin{equation}
    |\psi_\mathrm{e}, \psi_\mathrm{ph}\rangle \propto |E\rangle|0\rangle + g_{\mathrm{qu}} |E - \hbar\omega\rangle|1\rangle+\mathcal{O}(g_{\mathrm{qu}}^2).
\end{equation}

Our measurements are designed to probe this state by detecting single photons in coincidence with inelastically scattered electrons. In the experiment, a continuous electron beam at 120-keV energy is traversing a photonic chip based microresonator in an aloof geometry (Fig.~\ref{fig_setup}b).
The $\text{Si}_3\text{N}_4$ microresonator, fabricated using the photonic damascene process~\cite{Pfeiffer2016, Liu2021, Henke2021}, is designed for low optical loss, efficient fiber coupling and free-space near-field access (for optical characteristics, see Methods).
The narrow cross-section of the resonator (\SI{2.1}{\micro m}~$\times$~\SI{650}{nm}) is chosen such as to achieve electron-light velocity phase matching, and thereby enhanced coupling to the dielectric structure~\cite{Kozak2017_acceleration, Sapra2020, Dahan2020, Dahan2021, Kfir2020, Henke2021}, with predicted total photon generation probabilities of up to about 10\%. This results in a population of the empty cavity, with a peak emission around 0.8-eV photon energy (corresponding to an optical wavelength of \SI{1.5}{\micro m}). The energy and arrival time of each electron is measured with an event-based detector behind a magnetic prism spectrometer (Fig.~\ref{fig_setup}c). Cavity photons generated in the scattering process are coupled out to a bus waveguide, and further guided by optical fibers to a single-photon avalanche diode (SPAD). Figure \ref{fig_setup}d illustrates the preparation and detection of the electron-photon pair state via scattering at the resonator, followed by electron energy projection and coincidence measurement.

\section*{Spatial and spectral mapping of electron-induced cavity excitation}
\begin{figure} [!t]
\centering
\includegraphics[width=0.5\textwidth]{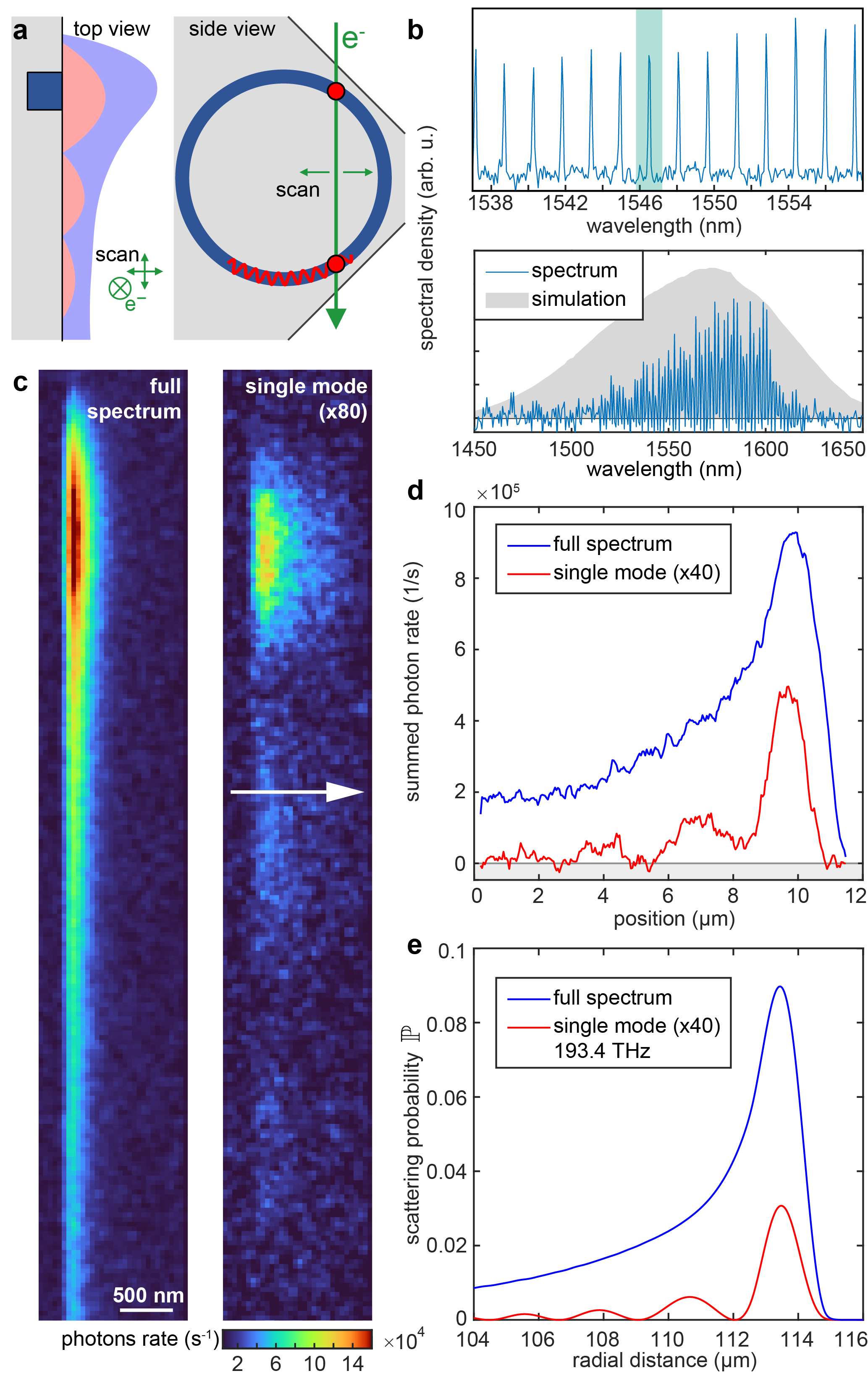}
\caption{\textbf{Spatial and spectral mapping of intracavity photon generation.}
\textbf{a)} Schematic of the measurement geometry. Top and side views of the photonic chip based microresonator (dark blue) embedded in the chip oxide layer (grey) with the electron beam (green) passing parallel to its surface, interacting twice with the resonator ring (red dots). The intensity patterns shown in panel \textbf{c} are indicated  by red and blue regions in the top view. \textbf{b)} Optical emission spectrum with electron beam centered above the microresonator waveguide (green area: single mode selected in panel \textbf{c}; grey area: simulated spectral envelope, see SI). \textbf{c)} STEM maps of the photon count rate (saturation-corrected, see Methods) above the resonator for the full spectrum (left, wavelengths 1520-1620 nm) and after spectral filtering to a single cavity mode (right, fiber Bragg grating at 1546.5-nm center wavelength). \textbf{d)} Signals from panel \textbf{c} integrated perpendicularly to the chip surface (white arrow) for full spectrum (blue) and single mode (red). \textbf{e)} Simulation of the position-dependent electron scattering probability (50-nm distance, see SI). }\label{fig_CL}
\end{figure}

We position the sample inside a transmission electron microscope (TEM), focus the low-convergence electron beam (25-nm diameter) in proximity of the photonic chip based resonator (cf. Fig.~\ref{fig_CL}a), and detect outcoupled photons, i.e., the cathodoluminescence. Measured with an optical spectrometer, the spectral density of the emission exhibits a comb-like structure (cf. Fig.~\ref{fig_CL}b), as a  result of free-electron coupling to the microresonator modes $a_\mu$ ($\mu$: mode index). The overall scattering probability {$\mathbb{P}=\sum _\mu |g_{{\mathrm{qu}},\mu}|^2$} is derived from the individual mode contributions $g_{{\mathrm{qu}},\mu}$ (a multi-mode description of the interaction is found in the SI). The spacing of the emission peaks in wavelength by \SI{1.58}{nm} matches the optically-characterized quasi-TM free-spectral-range (FSR) of \SI{194}{GHz}, implying the spontaneous excitation of these initially empty cavity modes. The predominant coupling to the quasi-TM mode family follows from its considerable azimuthal electric field component along the electron trajectory~\cite{Henke2021}. In comparison, quasi-TE and higher-order spatial modes have a weaker coupling to the mode-selective bus waveguide. The overall spectral range of detected modes spans from 1520-1620~nm, and is limited by the bandwidth of electron-light phase matching, wavelength-dependent coupling to the bus waveguide, as well as the detector spectral bandwidth, in good agreement with numerical simulations (cf. Fig.\ref{fig_CL}b).

A spatial characterization of the intracavity excitation is obtained by raster-scanning the electron beam near the waveguide using scanning transmission electron micoscopy (STEM) (Fig.~\ref{fig_CL}a, dark blue ring/box) and detecting the emitted light with a SPAD. The total and mode-specific coupling strengths are obtained by recording the photon count rate without (Fig.~\ref{fig_CL}c, left panel) and with (right panel), respectively, a spectral filter that selects a single TM cavity mode (indicated in Fig.~\ref{fig_CL}b). Regions where the electrons directly impinge on the chip show no photon signal, creating a sharp edge in the recorded 2D map. The strongest coupling and most efficient photon generation is observed for electrons passing the ring resonator tangentially, as expected for phase-matched free-electron light interaction. Both signals decay exponentially with distance from the structure, tracing the near-field mode profile in this spectral range. 

However, there are pronounced differences for the position dependence of the coupling strengths along the surface of the structure: the unfiltered count rate measured by the photon detector shows one prominent peak of the intracavity excitation and a smooth decay towards the resonator ring center, whereas the single-mode photon rate exhibits oscillations along the chip edge.
The oscillatory behavior of the single-mode coupling along the chip surface is caused by Ramsey-type interference~\cite{Echternkamp2016} between phase-mismatched sequential interactions of an electron (cf. Fig.~\ref{fig_CL}a, red dots) with the cavity vacuum field, as also observed for single-mode excitation of an externally-pumped ring resonators ~\cite{Henke2021}. Figure~\ref{fig_CL}d plots the position-dependent count rates along the chip surface, which are in good agreement with numerical simulations of the scattering probability {$\mathbb{P}$} (Fig.~\ref{fig_CL}e, see SI for details). Note that such interferences from sequential interactions or multipath scattering are characteristic of coherent cathodoluminescence~\cite{Talebi2014,Lingstadt2020,Schilder2020}.

\section*{Time- and energy-correlated electron-photon pairs}

The electron-driven spontaneous generation of photons in the photonic microresonator is expected to satisfy energy-momentum conservation~\cite{Bendana2011}. Therefore, each individual scattering event should change the energy of an electron by $-\hbar\omega$ and transfer the corresponding momentum to a cavity photon in the forward-propagating mode (clockwise around the resonator in Fig.~\ref{fig_CL}a). Investigating this process on the single-electron and single-photon level, we conduct event-based electron spectroscopy and time-tagged photon counting simultaneously. To this end, the electron beam is held fixed in the near field of the cavity, at a $\sim160$-nm distance from the surface. At this position, we detect photons with a probability of \SI{4.6e-5}{} per electron passing the structure. Considering coupling and detection losses (see Methods), this corresponds to an intrinsic generation probability of about $\sim 0.025$.

\begin{figure} [!t]
\centering
\includegraphics[width=0.5\textwidth]{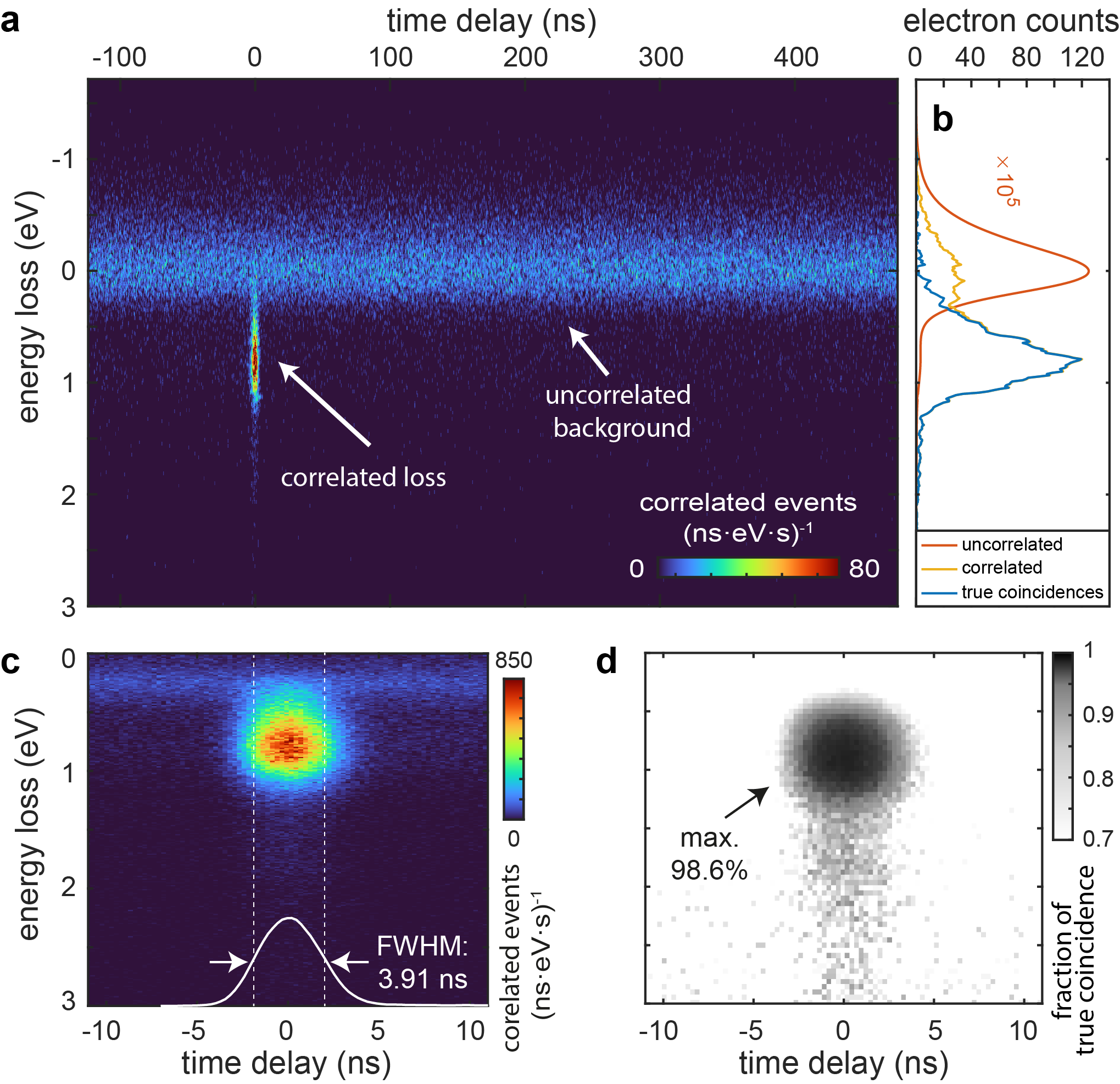}
\caption{\textbf{Photon-correlated electron energy-loss events.} \textbf{a)} Two-dimensional coincidence histogram as a function of the electron energy and time delay relative to a detected photon (30-s integration, ~0.65-pA electron current on detector). Electron-photon scattering events are evidenced by a pronounced peak around zero delay and 0.8~eV energy loss, separated from an uncorrelated background at 0~eV. \textbf{b)} Time-averaged spectral distributions of uncorrelated (red) and photon-correlated (yellow) electrons, as well as correlated events corrected for false coincidences (blue, see SI). \textbf{c)} Close-up of coincidence histogram (ZLP blocked, 60-s integration, 46-pA beam current at sample). White curve: Coincidence peak timing FWHM of 3.91~ns. \textbf{d)} Fraction of true coincidences of the correlation map in panel \textbf{c}, demonstrating the predominance of coincidence events over a low-noise background.}
\label{fig_correlations}
\end{figure}

The arrival time and kinetic energy of each electron is measured by event-based detection, using the stream of photons recorded by the SPAD as a time-tagging signal (cf. Fig.~\ref{fig_setup}a and Methods for details). Figure~\ref{fig_correlations}a shows the energy- and time-dependent histogram of electron arrivals relative to the photon detection event closest in time. The two main features observed are a time-independent background of accidental coincidences, and a sharp peak around 0.8-eV energy loss and 0-ns time delay (electronic and propagation delays subtracted, see Methods). The spectral distribution of the correlation peak (blue, yellow) and the uncorrelated signal (red), integrated over appropriate time intervals, are shown in Fig.~\ref{fig_correlations}b.
The correlated electron spectrum is downshifted by one photon energy, but otherwise closely matches the zero-loss-peak (ZLP) spectrum of the unscattered electrons in broadening and shape ($\sim \SI{0.5}{eV}$ width). This is consistent with the simulated and measured comparatively narrow electron-light phase matching bandwidth ($\sim 50$~meV, cf. Fig.~\ref{fig_CL}b).

Figure~\ref{fig_correlations}c displays the energy-time-histogram using a higher electron flux and the zero-loss peak largely eliminated with a mechanical slit in the electron spectrometer to prevent detector saturation. The background-corrected time profile (Fig.~\ref{fig_correlations}c inset) shows the precise temporal structure of the correlated electron-photon pairs with a width of 3.91~ns (FWHM). In principle, the temporal correlation should reproduce the cavity decay time, but it is not resolved in the present experiments (current device: $\sim$0.5-ns lifetime for a quality factor of $Q\sim5.5\times 10^5$; see Methods).

By selecting loss-scattered electrons within a specific time window, we identify correlated events on a single-particle basis. This enables inter-particle heralding schemes for either electrons or photons, quantified in terms of the measured rates of electrons ($R_\mathrm{e}$), photons ($R_\mathrm{p}$) and correlated events ($R_\mathrm{pe}$). The Klyshko heralding efficiencies $\eta_\mathrm{K}^{i}={R_\mathrm{pe}}/{R_{j}}$ ($i,j$=$\mathrm{e,p}$, $i\neq j$) describe the conditional probability of actually detecting a heralded particle~\cite{Signorini2020}. For the data shown in Fig.~3c, we measure $\eta_\mathrm{K}^\mathrm{p}\sim0.11\%$ and $\eta_\mathrm{K}^\mathrm{e}\sim57\%$ for photons and electrons, respectively.
The much more efficient heralding of electrons follows from considerably higher losses in the output coupling and detection of photons (cf. Methods).
Not being a fundamental physical limitation, we expect significant improvements on the photon collection efficiency with technical optimizations, including use of superconducting detectors and strongly over-coupled resonators. Taking into account particle losses in transmission and detection, we estimate intrinsic heralding efficiencies $\eta_\mathrm{I}^{i}$ of approximately 50\%. More generally, the phase-matched coupling to a specific mode family, and the aloof beam geometry that avoids undesired materials excitations, promise heralding efficiencies near unity (see detailed estimate of $\eta_\mathrm{I}^{i}$ in Methods \& SI).

We note that the measurement of the electron energy for each single event, in contrast to conventional optical spontaneous parametric down conversion, presents a direct measure of the energy quanta deposited in the optical cavity. In conjunction with the narrow phase-matching bandwidth, the event-based electron energy detection conditioned on the first energy loss sideband therefore represents an optical state projection onto a (non-classical) single-photon intracavity Fock state. Interactions with multiple electrons---relevant for studying electron-electron correlations~\cite{Kiesel2002, Kuwahara2021, Keramati2021}---can be excluded considering the multi-hit capability of the detector.

\section*{Photonic mode imaging using correlated electron-photon pairs}

Harnessing correlations of electrons with visible~\cite{Graham1986} or x-ray~\cite{Jannis2019} emission shows promise for enhancing contrast and resolution in electron spectroscopy~\cite{Rotunno2021}, with a wide range of applications in the study of core-level~\cite{Jannis2021} and very recently valence electronic excitations in nanomaterials~\cite{Varkentina2022}. Correlation-enhanced measurements isolate physical scattering events from uncorrelated noise such as detector dark counts. As a figure of merit, the coincidence-to-accidental ratio $\mathrm{CAR}=(R_\mathrm{pe}-R_\mathrm{acc})/R_\mathrm{acc}$ describes the rate of coincidence events $R_\mathrm{pe}$ over the uncorrelated background $R_\mathrm{acc}$ \cite{Signorini2020}. Figure \ref{fig_correlations}d displays the fraction of true coincidences, $(1-1/\mathrm{CAR})$ as a function of selected time delay and energy loss, reaching $98.6\%$, i.e., a $\mathrm{CAR}\sim75$. This demonstrates the high-fidelity generation of correlated electron-photon pairs for enhanced imaging.

\begin{figure} [!t]
\centering
\includegraphics[width=0.5\textwidth]{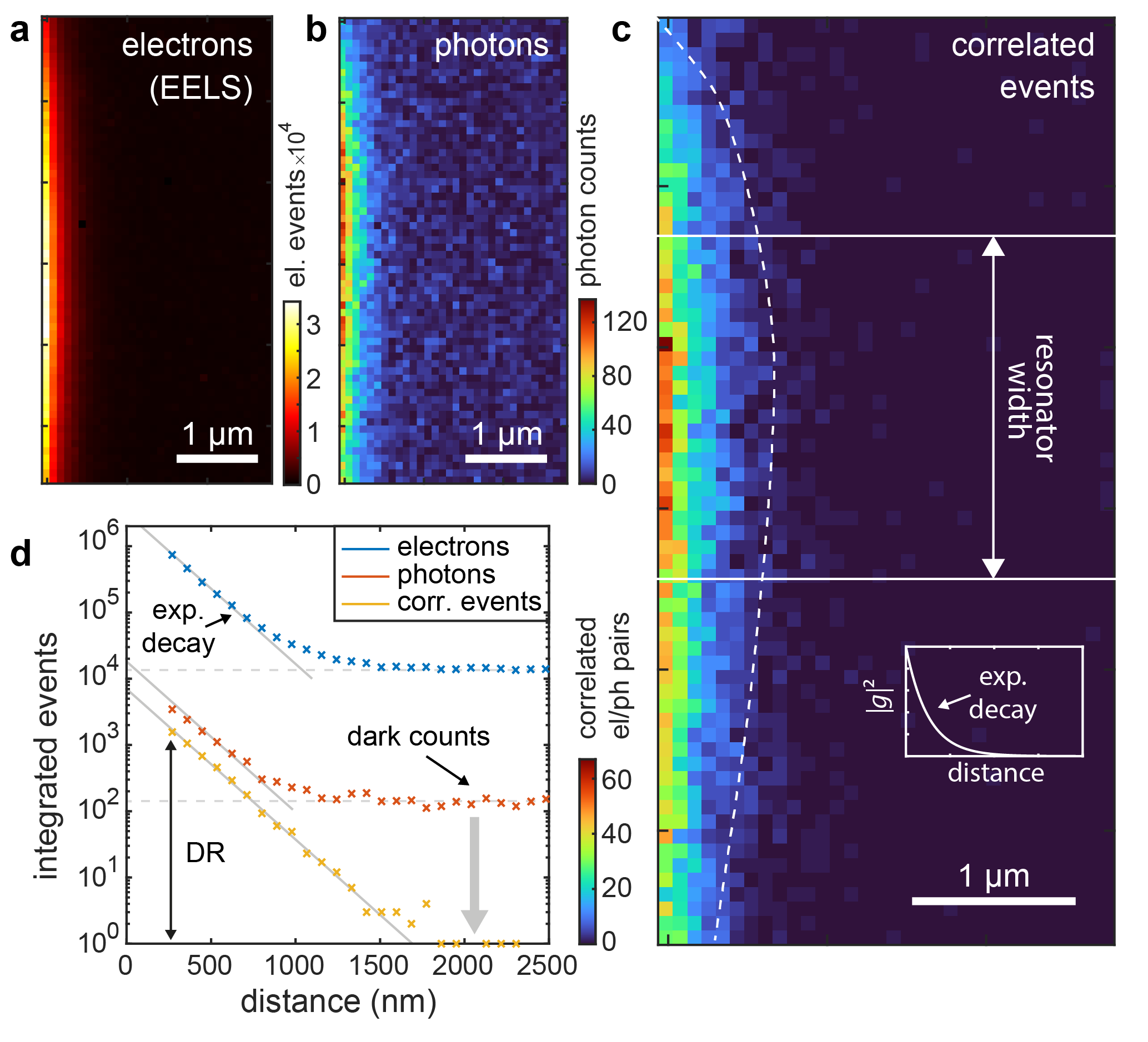}
\caption{\textbf{Imaging optical modes using time-correlated electron-photon pairs.} 
Spatial distribution of electron-photon correlations induced at the resonator mode with 30-ms integration time per pixel and starting around 270~nm above the chip surface (for geometry, see Fig.~\ref{fig_CL}): \textbf{a)} EELS, \textbf{b)} photon channel. \textbf{c)} Map of time-correlated electron-photon pairs. The envelope of the mode and position of the waveguide are indicated in white (dashed line serves as a guide to the eye, inset: sketch of the mode decay into the vacuum with scattering probability $|g|^2$). The energy-loss windows applied to EELS and correlated data are optimized over the imaging contrast individually.
\textbf{d)} Logarithmic plot comparing the counts of a)-c) as a function of distance to the surface, showing an exponential decay (integrated in area above the waveguide, in total, 1-s acquisition per point). The lowest noise and highest dynamic range ($>3$-orders of magnitude) is observed for the time-correlated imaging.}\label{fig_mapping}
\end{figure}

We provide a proof-of-concept demonstration by coincidence-gated raster-mapping of the resonator mode. Specifically, Figs.~\ref{fig_mapping}a,b show the time-integrated electron and photon signals, respectively, and Fig.~\ref{fig_mapping}c displays the correlated events only.
To quantify the correlation-induced improvement in image contrast, Figure~\ref{fig_mapping}d compares the respective count rates for the individual and correlated signals on a logarithmic scale. Both the photon and electron signals trace the exponential decay of the evanescent field away from the structure, leveling off at constant values for larger distances. These background offsets are determined by the uncorrelated noise rates, i.e., the photodetector dark counts (130~cts/s) and residual overlap of the ZLP with the energetic gate (\SI{1.5e4}{cts/s}), respectively.
In comparison, the correlated signal (cf. Fig.~\ref{fig_mapping}d, yellow curve) displays a reduced count rate, resulting from imperfect Klyshko heralding efficiencies $\eta_K^{i}<1$. However, the evanescent decay is resolved over longer distances, with the uncorrelated background noise suppressed by the narrow correlation window. As a result, the dynamic range (DR) is improved by two orders of magnitude, limited by the vanishing counts at large distances for the given integration time. (For a detailed analysis of the dynamic ranges and their relative enhancements, see the SI.)

\section*{Conclusions}
Our findings demonstrate and apply the high-fidelity generation and detection of correlated electron-photon pairs. Besides showing the capability for contrast enhancement in correlation-gated imaging, we implement flexible on-chip electron-heralded photon and photon-heralded single-electron sources with unique features and, in principle, near-unity heralding efficiency.
The integrated photonics platform allows for flexible electron-light phase-matching and spectral tuning of the generated cavity state. Post-selection of multi-loss electron events will facilitate the generation of tailored and higher-order photon Fock states. By merging free-electron quantum optics with integrated photonics, we provide a pathway towards a new class of hybrid quantum technology relying on entangled electrons and photons, with applications ranging from photonic quantum state synthesis to quantum-enhanced sensing and imaging.

\clearpage

\part*{Methods}
\begin{footnotesize}

\section{Simulation of inelastic electron scattering}

Assuming the electron beam trajectory is along the $\hat{z}$ direction with the transverse coordinates $(x,y)$, the mode-specific vacuum coupling strength between the electron and photon can be expressed as $g_{\mathrm{qu},\mu}(x,y)=\frac{e}{\sqrt{2\epsilon_\mu\hbar\omega_\mu V_\mu}}\int dz e^{iz\omega_\mu/v_e}u_\mu^*(x,y,z)$ (quantization procedure, see SI). Here, $e$ is the electron charge, $v_e$ the electron group velocity, 
$\epsilon_\mu$ the optical mode permittivity, $\omega_\mu$ the angular frequency, $V_\mu$ the mode volume, and $u_\mu(x,y,z)$ the $\hat{z}$ projection of the normalized electric field function along the electron trajectory. 
The factor $\int dz e^{iz\omega_\mu/v_e}u_\mu^*(z)$ reflects the phase-matching condition, and defines the phase-matching bandwidth 
dependent on the mode-specific dispersion of the resonator's effective refractive index.
Using our current resonator design with a small waveguide cross-section and a long interaction length of $\sim$\SI{40}{\micro m} (\SI{200}{fs} interaction time), the single-mode coupling strength is numerically predicted to be as high as $|g_{\mathrm{qu},\mu}|\sim0.03$ (equivalent coupling rate $|g_{{0},\mu}|/2\pi\sim\SI{2e10}{Hz}$) with an impact parameter of \SI{50}{nm} from the resonator surface, consistent with the experimental results.

For numerical simulations of the single mode coupling strength, we obtain the modal frequencies and field profiles via finite-element-method electromagnetic simulations (COMSOL Multiphysics). The vacuum coupling parameter for cavity modes $\omega_\mu$ is then calculated using: $g_{\mathrm{qu},\mu}(x,y) = \frac{\mathrm{e}}{2\hbar\omega_\mu\sqrt{n_{\mathrm{ph}}^\mu}}\int dze^{iz\omega_\mu/v_e}E_z^*(x,y,z)$, where $E_z$ is the $z$-directional complex electric field obtained from electromagnetic simulations~\cite{Henke2021}, and $n_{\mathrm{ph}}^\mu$ is the intracavity photon number inferred from power calculations. For the calculation of the photon generation (CL) and corresponding inelastic electron energy loss (EELS) probabilities, we assume a single-scattering electron-photon interaction with a weak interaction strength for individual modes $|g_{\mathrm{qu},\mu}|\ll1$, consistent with our experimental conditions. The Poissonian process is then simplified to: $P(|0_\mu\rangle) = \exp\left( -|g_{\mathrm{qu},\mu}|^2 \right) \approx 1 - |g_{\mathrm{qu},\mu}|^2$ and $P(|1_\mu\rangle) = \exp\left( -|g_{\mathrm{qu},\mu}|^2 \right)|g_{\mathrm{qu},\mu}|^2 \approx |g_{\mathrm{qu},\mu}|^2$. For EELS, the electron energy-loss probability for emitting a photon in the $\omega_\mu$ mode is $|g_{\mathrm{qu},\mu}|^2$. For CL, the photon ($\omega_\mu$-mode) detection probability is $|g_{\mathrm{qu},\mu}|^2\eta(\omega_\mu)S(\omega_\mu)$, considering the efficiency and bandwidth of the bus-waveguide-microresonator coupling $\eta(\omega)$ and the single-photon avalanche diode (SPAD) spectral sensitivity $S(\omega)$. Fiber optics loss is not considered at this point as it is frequency insensitive over the phase matching bandwidth. The SPAD response is taken from vendor specifications, and the bus-waveguide-microresonator coupling efficiency is calculated from $\eta(\omega)=\kappa_{\mathrm{ex}}(\omega)/(\kappa_0(\omega)+\kappa_{\mathrm{ex}}(\omega))$ with experimentally measured external coupling rate $\kappa_{\mathrm{ex}}(\omega)$ and intrinsic loss rate $\kappa_0(\omega)$. More details can be found in the Supplementary Information.

\section{Fiber-integrated silicon nitride microresonators}

The Si$_3$N$_4$ microresonator (wafer D66\_01\_F1\_C20) was fabricated using the photonic Damascene process~\cite{Pfeiffer2016, Liu2021}. Si$_3$N$_4$ exhibits very low absorption losses in the telecom region ($<1$ dB/m for 1550nm) and can be engineered to achieve optimal dispersion, and therefore phase-matching conditions with the passing electrons, as well as good fiber coupling~\cite{Liu2017}. The used resonator structure with a \SI{2.1}{\mu m} $\times$ \SI{650}{nm} waveguide size and a ring radius of \SI{113.75}{\mu m} (194~GHz free spectral range for quasi-TM mode family) was designed for phase matching at an optical wavelength around 1550~nm and an electron energy of 120~keV. The mean quality factor ($Q$-factor) of the observed resonances is $\sim5.5\times 10^5$, with the resonator-waveguide coupling efficiency degraded to $\sim 17\%$ during the experiment, likely caused by an increased intrinsic loss from charging induced carbon deposition on the resonator surface. More details on the chip fabrication can be found in~\cite{Henke2021}.

The chip is transferred into the TEM via a custom made holder with a hollowed-out tube and a KF blind flange with vacuum fiber feedthroughs at the end. This allows fibers connected to the chip at the holder tip to be fed through the holder to the feedthroughs and through them be connected to the outside of the TEM. The adapter piece and T-shaped base plate at the tip of the holder are designed to place the chip into the rotation axis of the holder, allowing for precise chip rotation in the sample plane of the electron microscope.

The bus waveguide on the photonic chip is connected to ultra-high numerical aperture (UHNA-7) fibers (mode field diameter of $\sim \SI{3.2}{\mu m}$ at 1550nm), which are spliced to standard single-mode fibers (SMF-28) with a splicing loss of $<0.2$ dB~\cite{Raja2020}. The optical transmission of photons coupled into the bus waveguide to the output of the external fiber connector of the holder is measured to be $\sim 40\%$.

\section{(S)TEM instrumentation}
The experiments are conducted at the Göttingen UTEM instrument that is based on a Schottky field-emission TEM (JEOL JEM 2100F)~\cite{Feist2017}. The electron gun is operated with a continuous electron beam in the extended Schottky regime with an under-heated emitter, resulting in an energy spread of 0.5~eV at a beam energy of 120~keV. The overall beam current is adjusted by the emitter temperature and condenser aperture to prevent oversaturation of the detectors (Fig.~\ref{fig_correlations}a\&b 40-\si{\micro m} aperture, Fig.~\ref{fig_correlations}c\&d 100-\si{\micro m} aperture). Reducing beam clipping at the spatially extended structures, the low-magnification STEM mode (LM-STEM) is used in the experiments, achieving an electron focal spot sizes of $<25$~nm. The beam is held fixed above the sample surface (Fig.~\ref{fig_correlations}, e.g. 100-200~nm distance) or raster scanned (Fig.~\ref{fig_CL}\&\ref{fig_mapping}, 30-ms dwell time). The event-based spectral analysis of the transmitted electron beam is described below.

\section{Optical setup}
The photons generated at the mounted microresonator are guided to the detectors through standard single-mode patch cables (SMF-28).
We carried out an evaluation on existing photon loss channels between the generation and detection, including fiber transmission and detection efficiencies (table see SI), leading to a full transmission and detection probability of 0.3(2)\%.
The spectral detection bandwidth of the optical setup is mainly given by the wavelength-dependent coupling efficiency between the resonator and the bus waveguide as well as the detection bandwidth of the detectors.

The optical spectra shown in Fig.~\ref{fig_CL}b are measured using a Czerny-Turner spectrometer (Horiba iHR550, 600~l/mm grating, 50~\si{\micro m} slit width, 0.4~nm estimated spectral resolution) with a fiber adapter and a liquid nitrogen cooled, NIR sensitive InGaAs camera (Horiba Symphony II, 512$\times$1 pixel, {50}-\si{\micro m} pixel size, 800-1650~nm detection bandwidth). Spectral calibration was performed with a strongly attenuated fiber-coupled and tunable continuous-wave (cw) laser (Toptica, CTL 1550).

All further experiments are conducted with a single-photon avalanche diode (SPAD ID230, IDQuantique SA, 900-1700 nm full detection bandwidth) operated in Geiger mode and cooled to $-90^\circ$C. The detector is set to a detection efficiency of $>25\%$ (1560~nm wavelength, specified by the manufacturer). Adjusting for the overall photon count rate, the detector dead time is set either to 10 or 50~\si{\micro s}, resulting in dark count rates of 2000~cts/s or 130~cts/s, respectively, reaching the specified intrinsic dark count rates. At high count rates, detector saturation is corrected for the data shown in Fig.~\ref{fig_CL} (see SI). The photon arrival at the SPAD is translated to a LV-TTL signal ($\approx 150$-ps resolution, 100-ns pulse width) and passed to the event-based electron detector for further counting and correlation.

For optical mode filtering, the reflection from a manually tunable fiber Bragg grating (FBG, Advanced Optics Solutions GmbH, 1540-1550~nm tuning range, bandwidth FWHM $\sim$100~GHz ) is employed (along with a circulator). The FBG leads to an overall loss of about 50\% (see SI). For pre-alignment of the FBG to a resonance, we connect the cw laser to the other side of the resonator bus waveguide while observing the transmission signal on a photodiode. Precise tuning is done by optimizing the SPAD signal, excluding potential thermal drifts of the resonance wavelength induced by the laser. We note, that the ratio of the single- and multi-mode count rates at the resonator position (cf. Fig.~\ref{fig_CL}d) indicates that about 100 modes contribute to the unfiltered map shown in Fig.~\ref{fig_CL}c.

\section{Event-based electron analysis}
The experiments measuring single particles are performed using a commercial hybrid pixel electron detector based on the Timepix3 ASIC (EM CheeTah T3, Amsterdam Scientific Instruments B.V.) mounted behind an energy-dispersive spectrometer (CEFID, CEOS GmbH). The camera generates a stream of data packages containing the position and timing of electron-activated detector pixels digitized with 1.56-ns time bins. Additional data packages are generated from the internally synchronized LV-TTL input at one of two Time-to-Digital-Converters (TDCs) of the SPIDR readout system with $\sim$260-ps timing precision. Hereby, an external synchronization of the electron arrival time and photon counts at the SPAD becomes possible using the joint global time stamp of the detector and TDC channels. Technologically, event-based imaging in electron microscopy enables new experimental concept, including background suppression in X-ray spectroscopy~\cite{Jannis2019}, continuous illumination picosecond imaging~\cite{Wessels2021} and dose-optimized high-speed STEM~\cite{Jannis2022, Auad2021}.

Each image pixel shows a fixed timing offset relative to the mean detector response, that must be determined for precise measurement of the time of arrival (ToA) of the electrons~\cite{Pitters2019}. This per-pixel calibration is performed with a pulsed electron beam generated by photoemission using a femtosecond laser amplifier (Pharos, Light Conversion). Considering the sub-picosecond electron pulses generated~\cite{Feist2017} and a $<$500-ps timing jitter of the laser trigger output, a sub-ns timing precision is achieved. Specifically, the average offset in time of arrival is determined for integrating ~1750 hits per pixel, showing a spread of up to 8~ns in the raw data over the entire four-quadrant detector (512$\times$512 pixel) that can be corrected for.

In the analysis of electron-photon correlations, all electron-generated detector hits are tagged with their relative timing to the closest TDC data packet from the SPAD output. For the different experiments, the dead time of the SPAD was chosen between 10~$\mu$s and 50$~\mu$s and only hits within 600-ns relative timing are taken for the further data analysis. Therefore, all considered electrons can be uniquely assigned to specific photon counts of the SPAD.

Single-particle clustering using the $k$-means++ clusters algorithm (implemented in MATLAB 2020b, MathWorks Inc.) is applied on the stream of detector hits to form event clusters, each representing a single electron. At 120-keV beam energy, the maximum distance of hits in a cluster is set to 5 pixels or 125~ns, preventing mixing of events and maintaining the multi-hit capability of the detector. Event localization is performed by retrieving the arithmetic mean position and the earliest arrival time of all hits within a cluster. The overall electron current on the detector is estimated by the number of hits per time and typical cluster-size distribution at 120~kev (on average 4 hits per cluster, detector saturation at 120~MHit/s or ~4.8~pA for homogeneous illumination).

Finally, a 2D histogram of the electron energy and the relative TDC timings results in the plots shown in Figure~\ref{fig_correlations}a\&b. The data is binned in time units of the TDC channels of $\sim$~260~ps (1/3.84~GHz) and the spectral pixel size of 0.11~eV/px. The overall temporal width of the correlation peak of 3.91~ns (FWHM, see Fig.~\ref{fig_correlations}c) is mostly determined by strong variations in cluster size and shape for high-energy electrons and might be improved in future studies using advanced clustering methods considering real particle traces (cf. Ref.~\cite{vanSchayck2020}). The absolute electronic and propagation delays are subtracted for clarity and the data is centered in time around the peak of correlated events. Residual 50-Hz and further uncorrelated timing jitter in the experiment is corrected in the energy disperse axis by considering the uncorrelated electron background with 100-\si{\micro s} time bins.

\section{Heralding Efficiencies}
The efficiency of the heralding scheme can be estimated using the Klyshko heralding efficiency (also called total heralding efficiency) $\eta_\mathrm{K}^{i}={R_\mathrm{pe}}/{R_{j}}$ ($i,j$=$\mathrm{e,p}$, $i\neq j$) and the intrinsic heralding efficiency $\eta_\mathrm{I}^{i}=\eta_\mathrm{K}^{i}/(\eta_\mathrm{D}^{i}T_{i}) $~\cite{Signorini2020}, with the measured rates of electrons $R_e$, photons $R_p$ and correlated events $R_{pe}$. While the intrinsic heralding efficiency describes the probability of a pair generation source, in our case the resonator as electron-cavity interaction region, to produce a particle pair, the Klyshko efficiency adds the particle losses due to the setup. The various sources of photon loss in the measurement setup, consisting of coupling efficiency from the resonator to the bus waveguide, various fiber coupling and propagation losses, SPAD detection efficiency, saturation losses and overlap with the expected photon bandwidth, as well as losses due to photons scattered into the backwards-propagating direction, were analysed individually, taking into account the respective error margins (see SI). On the electron side, the quantum efficiency of the event-based electron detector is calibrated by measuring the electron beam current at the entrance of the EELS spectrometer. Further transmission losses of electrons may include partial clipping of the beam at the photonic chip itself, or at apertures along the way to the detector. These were not included in the analysis since no precise current measurement is available at the sample position, which may lead to an underestimation of the intrinsic heralding efficiency $\eta_\mathrm{I}^\mathrm{e}$.
We estimate an intrinsic photon heralding efficiency $\eta_\mathrm{I}^\mathrm{p}$ reaching 49(21)\%, with larger uncertainties to be studied in future experiments (see SI). Considering known loss channels, for heralding single electrons, we experimentally retrieve an intrinsic heralding efficiency $\eta_\mathrm{I}^\mathrm{e}$ above 63(10)\%.

\section*{Data Availability Statement} The code and data used to produce the plots within this work will be released on the repository \texttt{Zenodo} upon publication of this preprint.

\section*{Acknowledgments} 
We thank R. Haindl and E. Maddox for discussions on the timing calibration and event clustering. We thank M. M\"oller for the support on the STEM alignment. We thank J. Borchert and T. Weitz for providing us with the NIR liquid-nitrogen cooled spectrometer. We thank members of the eBEAM consortium, in particular A. Polman, F.~J. García de Abajo, N. Talebi, J. Verbeeck, and M. Kociak, for useful discussions and feedback.

\noindent\textbf{Funding Information:}
All samples were fabricated in the Center of MicroNanoTechnology (CMi) at EPFL. This material is based upon work supported by the Air Force Office of Scientific Research under award number FA9550-19-1-0250.  This work was further supported by the Swiss National Science Foundation under grant agreements 185870 (Ambizione), 182103 and 176563 (BRIDGE). 
The work at the Göttingen UTEM Lab was funded by the Deutsche Forschungsgemeinschaft (DFG, German Research Foundation) through  432680300/SFB\,1456 (project C01) and the Gottfried Wilhelm Leibniz program, and the European Union’s Horizon 2020 research and innovation programme under grant agreement No. 101017720 (FET-Proactive EBEAM). Y.Y. acknowledges support from the EU H2020 research and innovation program under the Marie Sklodowska-Curie IF grant agreement No. 101033593 (SEPhIM). O.K. acknowledges the Max Planck Society for funding from the Manfred Eigen Fellowship for postdoctoral fellows from abroad.\\

\noindent\textbf{Author contribution:}
A.S.R., G.H. and Y.Y. designed the photonic chip devices, supported by O.K. and J.L..
J.L., Z.Q. and R.N.W. developed the fabrication process and fabricated the devices.
A.S.R. optically characterized and packaged the devices, supported by Y.Y. and G.H..
J.W.H. designed the TEM sample mount.
G.A. build the optical setup supported by A.F., J.W.H. and F.J.K..
A.F. and G.A. performed the TEM experiments supported by J.W.H. and F.J.K..
A.F. implemented the clustering and time-tagging of the event-based data.
The data was analysed by A.F. and G.A. supported by G.H. and H.L.M..
Y.Y. and G.H. performed the numerical simulations supported by A.F., G.A. and O.K..
G.H. and C.R. devised the theory section with support from A.F..
The study was planned and directed by C.R. and T.J.K.
The manuscript was written by A.F., G.H., G.A., Y.Y., T.J.K., and C.R., after discussions with and input from all authors.\\

\noindent\textbf{Additional information:}
Correspondence and requests for materials should be addressed to C.R. and T.J.K. \\
(\texttt{claus.ropers@mpinat.mpg.de, tobias.kippenberg@epfl.ch})\\

\noindent\textbf{Competing financial interests:}
The authors declare no competing financial interests.\\

\end{footnotesize}

\bibliographystyle{apsrev4-2}
\bibliography{manuscript_CL}

\end{document}